\documentclass[aps,floatfix,prd,twocolumn,showkeys]{revtex4}
\usepackage{acronym,dcolumn,epstopdf,graphicx,hyperref,ragged2e,textcomp}

\usepackage[usenames]{color}
\usepackage[normalem]{ulem}

\begin{document}
\title{Directed searches for continuous gravitational waves from twelve
supernova remnants in data from Advanced LIGO's second observing run}
\author{Lee Lindblom}
\affiliation{
  Center for Astrophysics and Space Sciences,
  University of California at San Diego,
  La Jolla, California 92093-0424
}
\author{Benjamin J. Owen}
\affiliation{
Department of Physics and Astronomy,
Texas Tech University,
Lubbock, Texas 79409-1051,
USA
}

\begin{abstract}
We describe directed searches for continuous gravitational waves from twelve
well localized non-pulsing candidate neutron stars in young supernova remnants
using data from Advanced LIGO's second observing run.  We assumed that each
neutron star is isolated and searched a band of frequencies from
15 to 150\,Hz, consistent with frequencies expected from known young pulsars.
After coherently integrating spans of data ranging from 12.0 to 55.9 days using
the $\mathcal{F}$-statistic and applying data-based vetoes, we found no evidence
of astrophysical signals.  We set upper limits on intrinsic gravitational wave
amplitude in some cases stronger than $10^{-25},$ generally about a factor of
two better than upper limits on the same objects from Advanced LIGO's first
observing run.
\end{abstract}


\maketitle

\acrodef{GW}{gravitational wave}
\acrodef{O1}{first observing run}
\acrodef{O2}{second observing run}
\acrodef{SFT}{short Fourier transform}
\acrodef{SNR}{supernova remnant}

\section{Introduction}

Young isolated neutron stars and suspected locations of the same are promising
targets for directed searches for continuous \acp{GW} \cite{Glampedakis2018}.
Even without timing obtained from electromagnetic observations of a pulsar, such
searches can achieve interesting sensitivities for reasonable computational
costs~\cite{Wette2008}.
Young \acp{SNR} containing candidate non-pulsing neutron stars are natural
targets for such searches, as are small \acp{SNR} or pulsar wind nebulae even in
the absence of a candidate neutron star (as long as the \ac{SNR} is not
Type~Ia, which does not leave behind a compact object).

Many upper limits on continuous \acp{GW} from isolated, well localized neutron
stars other than known pulsars have been published over the last decade.
These have used data ranging from Initial LIGO runs to Advanced LIGO's \ac{O1}
and \ac{O2}.
Most searches targeted relatively young \acp{SNR}~\cite{S5CasA, S5Stochastic,
S6NineSNRs, Sun2016, Zhu2016, O1Stochastic, O1FifteenSNRs, Ming2019,
O1O2Stochastic}.
Some searches targeted promising small areas such as the galactic
center~\cite{S5Stochastic, S5GalCen, O1Stochastic, O1O2Stochastic,
Piccinni2020}.
One search targeted a nearby globular cluster, where multi-body interactions
might effectively rejuvenate an old neutron star for purposes of continuous
\ac{GW} emission~\cite{S6Globular}.
Some searches used short coherence times and fast, computationally cheap methods
originally developed for the stochastic \ac{GW} background~\cite{S5Stochastic,
O1Stochastic, O1O2Stochastic}.
Most searches were slower but more sensitive, using longer coherence times and
methods specialized for continuous waves based on matched filtering and similar
techniques.

Here we present the first searches of \ac{O2} data for twelve \acp{SNR}, using
the fully coherent $\mathcal{F}$-statistic as implemented in a code pipeline
descended from the one used in the first published search~\cite{S5CasA} among
others~\cite{S6NineSNRs, O1FifteenSNRs}.
Since the \ac{O2} noise spectrum is not much lower than \ac{O1}, we deepened
these searches with respect to \ac{O1} searches~\cite{O1FifteenSNRs} by focusing
on low frequencies compatible with those observed in young
pulsars~\cite{Manchester2005}.
This focus allowed us to increase coherence times and obtain significant
improvements in sensitivity over \ac{O1}.
Low frequencies have the drawback, however, that greater neutron star
ellipticities or $r$-mode amplitudes are required to generate detectable
signals.
We did not search three \acp{SNR} from the list in Ref.~\cite{O1FifteenSNRs}
because the Einstein@Home distributed computing project has already searched
them~\cite{Ming2019} to a depth which cannot be matched without such great
computing resources which we do not have.
We also did not search Fomalhaut~b as in Ref.~\cite{O1FifteenSNRs} because our
code (although improved over previous versions) is inefficient for targets with
such long spin-down timescales.
In the future we plan to improve the code to efficiently search higher
frequencies and longer spin-down timescales.
For now our searches are interesting as the most sensitive yet (in strain) for
these twelve \acp{SNR}.

\section{Searches}

\begin{table*}
  \begin{center}
    \begin{tabular}{rclccD{.}{.}{2.1}cD{.}{.}{2.2}c}
      \hline
      \hline
      \multicolumn{1}{c}{SNR} & parameter & Other name & RA+dec & Ref.\ & \multicolumn{1}{c}{$D$} & Ref.\ & \multicolumn{1}{c}{$a$} & Ref.\ \\
      \multicolumn{1}{c}{(G name)} & space & & (J2000 h:m:s+d:m:s) & & \multicolumn{1}{c}{(kpc)} & & \multicolumn{1}{c}{(kyr)} & \\
      \tableline
1.9+0.3 & & --- & 17:48:46.9\textminus27:10:16 & \protect\cite{Reich1984} & 8.5 &
\protect\cite{Reynolds2008} & 0.1 & \protect\cite{Reynolds2008}
\\
15.9+0.2 & wide & --- & 18:18:52.1\textminus15:02:14 & \protect\cite{Reynolds2006} &
8.5 & \protect\cite{Reynolds2006} & 0.54 & \protect\cite{Reynolds2006}
\\
15.9+0.2 & deep & --- & 18:18:52.1\textminus15:02:14 & \protect\cite{Reynolds2006} &
8.5 & \protect\cite{Reynolds2006} & 2.4 & \protect\cite{Reynolds2006}
\\
18.9\textminus1.1 & & --- & 18:29:13.1\textminus12:51:13 &
\protect\cite{Tullmann2010} & 2 & \protect\cite{Harrus2004} & 4.4 &
\protect\cite{Harrus2004}
\\
39.2\textminus0.3 & & 3C 396 & 19:04:04.7+05:27:12 & \protect\cite{Olbert2003} & 6.2
& \protect\cite{Su2011} & 3 & \protect\cite{Su2011}
\\
65.7+1.2 & & DA 495 & 19:52:17.0+29:25:53 & \protect\cite{Arzoumanian2008} & 1.5 &
\protect\cite{Kothes2004} & 20 & \protect\cite{Kothes2008}
\\
93.3+6.9 & & DA 530 & 20:52:14.0+55:17:22 & \protect\cite{Jiang2007} & 1.7 &
\protect\cite{Foster2003} & 5 & \protect\cite{Jiang2007}
\\
189.1+3.0 & wide & IC 443 & 06:17:05.3+22:21:27 & \protect\cite{Olbert2001} & 1.5 &
\protect\cite{Fesen1980} & 3 & \protect\cite{Petre1988}
\\
189.1+3.0 & deep & IC 443 & 06:17:05.3+22:21:27 & \protect\cite{Olbert2001} & 1.5 &
\protect\cite{Fesen1980} & 20 & \protect\cite{Swartz2015}
\\
291.0\textminus0.1 & & MSH 11\textminus62 & 11:11:48.6\textminus60:39:26 &
\protect\cite{Slane2012} & 3.5 & \protect\cite{Moffett2001} & 1.2 &
\protect\cite{Slane2012}
\\
330.2+1.0 & wide & --- & 16:01:03.1\textminus51:33:54 & \protect\cite{Park2006} & 5
& \protect\cite{McClure-Griffiths2001} & 1 & \protect\cite{Park2009}
\\
330.2+1.0 & deep & --- & 16:01:03.1\textminus51:33:54 & \protect\cite{Park2006} & 10
& \protect\cite{McClure-Griffiths2001} & 3 & \protect\cite{Torii2006}
\\
350.1\textminus0.3 & & --- & 17:20:54.5\textminus37:26:52 &
\protect\cite{Gaensler2008} & 4.5 & \protect\cite{Gaensler2008} & 0.6 &
\protect\cite{Lovchinsky2011}
\\
353.6\textminus0.7 & & --- & 17:32:03.3\textminus34:45:18 &
\protect\cite{Halpern2010} & 3.2 & \protect\cite{Tian2008} & 27 &
\protect\cite{Tian2008}
\\
354.4+0.0 & wide & --- & 17:31:27.5\textminus33:34:12 & \protect\cite{Roy2013} & 5 &
\protect\cite{Roy2013} & 0.1 & \protect\cite{Roy2013}
\\
354.4+0.0 & deep & --- & 17:31:27.5\textminus33:34:12 & \protect\cite{Roy2013} & 8 &
\protect\cite{Roy2013} & 0.5 & \protect\cite{Roy2013}
\\
\hline
\end{tabular}
\end{center}

\caption{
\label{t:table1}
Astronomical parameters of \acp{SNR} used in each search:
Right ascension and declination, distance $D$, age $a$, and references for each
parameter.
For \acp{SNR} whose range of age and distance estimates in the literature is not
too great, the search used the optimistic (nearby and young) end of the range.
For some \acp{SNR} the range was great enough to justify separate wide parameter
searches (optimistic) and deep parameter searches (pessimistic).
}
\end{table*}

In most respects the searches were done similarly to~\cite{O1FifteenSNRs}, so we
summarize briefly and refer the reader to that paper for further details.
The same goes for the upper limits described in the next Section.

\subsection{Setup}

We made the usual assumptions about the signals, that they had negligible
intrinsic amplitude evolution and that their frequency evolution in the frame of
the solar system barycenter was given by
\begin{equation}
\label{ft}
f(t) = f + \dot f (t-t_0) + \frac{1}{2} \ddot f (t-t_0)^2,
\end{equation}
where $t_0$ is the beginning of the observation, the frequency derivatives are
evaluated at that time, and we write a simple $f$ for $f(t_0).$
Hence our searches were sensitive to neutron stars without binary companions,
significant timing noise, or glitches; and spinning down on timescales much
longer than the duration of any observation.

We used the multi-detector $\mathcal{F}$-statistic~\cite{Jaranowski1998,
Cutler2005}, which combines matched filters for the above type of signal in such
a way as to account for amplitude and phase modulation due to the daily rotation
of the detectors with relatively little computational cost.
In stationary Gaussian noise, $2\mathcal{F}$ is drawn from a $\chi^2$
distribution with four degrees of freedom.
The $\chi^2$ is noncentral if a signal is present.
For loud signals the amplitude signal-to-noise ratio is roughly
$\sqrt{\mathcal{F}/2}.$

We used Advanced LIGO \ac{O2} data~\cite{Vallisneri2015, GWOSC} with version
\texttt{C02} calibration and cleaning as described in~\cite{Cahillane2018}.
Thus the amplitude calibration uncertainties were no greater than 8\% for each
interferometer.
As in previous searches of this type, we used strain data processed into
\acp{SFT} of 1800\,s duration, high pass filtered and Tukey windowed.
And we chose the set of \acp{SFT} for each search, once its time span was fixed
(see below), by minimizing the harmonic mean of the noise power spectral density
over the span and the frequency band.

With the direction to each candidate neutron star known, the parameter space of
each search was the set $(f, \dot f, \ddot f).$
In contrast to Ref.~\cite{O1FifteenSNRs} and earlier searches, we fixed
$f_{\min}$ and $f_{\max}$ at 15\,Hz and 150\,Hz respectively.
Our goal was to improve the sensitivity significantly over earlier \ac{O1}
results~\cite{O1FifteenSNRs, Ming2019}, even though the strain noise was only
slightly improved, while focusing on a range of frequencies compatible with the
emission expected from known young pulsars~\cite{Manchester2005}.
Rounding up a bit from the 124\,Hz expected from the fastest known young pulsar,
we set $f_{\max}$ to 150\,Hz.
Since the precise value of $f_{\min}$ has very little effect on the cost of the
searches, we somewhat arbitrarily set it to 15\,Hz where the noise spectrum is
rising steeply.
The ranges of frequency derivatives were then chosen as in~\cite{O1FifteenSNRs},
with
\begin{equation}
-\frac{f}{a} \le \dot f \le -\frac{1}{6} \frac{f}{a}
\end{equation}
for a given $f$ and
\begin{equation}
2\frac{\dot f^2}{f} \le \ddot f \le 7\frac{\dot f^2}{f}
\end{equation}
for a given $(f, \dot f).$
Thus we were open to a wide but physically motivated range of possible emission
scenarios.

\subsection{Target List}

\begin{table*}
\begin{center}
\begin{tabular}{rcccrrrrD{.}{.}{2.1}}
\hline
\hline
\multicolumn{1}{c}{SNR} & parameter & $T_\mathrm{span}$ & $T_\mathrm{span}$ &
\multicolumn{1}{c}{Start of span} & \multicolumn{1}{c}{H1} &
\multicolumn{1}{c}{L1} & \multicolumn{1}{c}{Duty} &
\multicolumn{1}{c}{$h_0^\mathrm{age}$}
\\
\multicolumn{1}{c}{(G name)} & space & (seconds) & (days) &
\multicolumn{1}{c}{(UTC, 2017)} & \multicolumn{1}{c}{SFTs} &
\multicolumn{1}{c}{SFTs} & \multicolumn{1}{c}{factor} &
\multicolumn{1}{c}{$(\times10^{-25})$}
\\
\tableline
1.9+0.3 &  & 1,036,229 & 12.0 & Jun 23 03:59:29 & 460 & 466 & 0.80 & 8.5
\\
15.9+0.2 & wide & 1,744,260 & 20.2 & Aug 04 21:11:52 & 753 & 748 & 0.77 & 3.6
\\
15.9+0.2 & deep & 2,593,109 & 30.0 & Jul 26 21:41:05 & 1076 & 1095 & 0.75 & 1.7
\\
18.9\textminus1.1 & & 3,014,418 & 34.9 & Jul 22 00:39:16 & 1204 & 1272 & 0.74
& 5.4
\\
39.2\textminus0.3 & & 2,734,846 & 31.7 & Jul 23 17:19:34 & 1106 & 1152 & 0.74
& 2.1
\\
65.7+1.2 & & 4,450,430  & 51.5 & Jan 19 08:03:58 & 1916 & 1580 & 0.71 & 3.4
\\
93.3+6.9 & & 3,067,958 & 35.5 & Jul 21 08:46:56 & 1224 & 1288 & 0.74 & 6.0
\\
189.1+3.0 & wide & 2,739,425 & 31.7 & Jul 23 16:03:15 & 1108 & 1154 & 0.74 & 8.8
\\
189.1+3.0 & deep & 4,468,104 & 51.7 & Jan 19 03:09:24 & 1917 & 1588 & 0.71 & 3.4
\\
291.0\textminus0.1 & & 2,160,350  & 25.0 & Jul 28 03:45:55 & 906 & 913 & 0.76
& 5.9
\\
330.2+1.0 & wide & 2,056,663 & 23.8 & Aug 02 00:41:51 & 876 & 865 & 0.73 & 4.6
\\
330.2+1.0 & deep & 2,765,446 & 32.0 & Jul 23 08:49:34 & 1116 & 1169 & 0.74 & 1.3
\\
350.1\textminus0.3 & & 1,794,825 & 20.8 & Aug 05 03:25:49 & 777 & 773 & 0.78 &
6.5
\\
353.6\textminus0.7 & & 4,827,338 & 55.9 & Jul 01 01:03:56 & 1581 & 1955 & 0.66
& 1.4
\\
354.4+0.0 & wide & 1,040,749 & 12.0 & Jun 23 02:59:29 & 462 & 469 & 0.81 & 14.4
\\
354.4+0.0 & deep & 1,694,450 & 19.6 & Aug 05 03:32:02 & 736 & 722 & 0.77 & 4.0
\\
\hline
\end{tabular}
\end{center}

\caption{
\label{t:table2}
Derived parameters used in each search.
The duty factor is the total \protect\ac{SFT} time divided by $T_\mathrm{span}$
divided by the number of interferometers (two).  As in the previous table, for
objects with two entries the first is a wide search (optimistic
parameter estimates) and the second is a deep search (pessimistic
parameter estimates).  The ranges used for the spin-down parameters
(described in the text) for wide and deep searches are not the same.
}
\end{table*}

Our choice of targets was based on the same criteria adopted in
the \ac{O1} search~\cite{O1FifteenSNRs}.  We required
that our search of a particular target at fixed computational cost be
sensitive enough to detect the strongest continuous \ac{GW} signal
consistent with conservation of energy.  This strongest
signal, based on the age $a$ and distance $D$ of the source,
\begin{equation}
h_0^\mathrm{age} = 1.26\times10^{-24} \left( \frac{\mbox{3.30 kpc}} {D}
\right) \left( \frac{\mbox{300 yr}} {a} \right)^{1/2},
\end{equation}
is analogous to the spin-down limit for known pulsars and indicates
the strongest possible intrinsic amplitude produced by an object whose
unknown spin-down is entirely due to \ac{GW} emission and has been
since birth~\cite{Wette2008}.  The intrinsic amplitude  $h_0$
characterizes the \ac{GW} metric perturbation without reference to any
particular orientation or polarization~\cite{Jaranowski1998}, and
therefore is typically a factor 2--3 larger than the actual strain reponse of
the interferometers.

As in the \ac{O1} search~\cite{O1FifteenSNRs} we selected targets from Green's
catalog of \acp{SNR} \cite{Green2019} (now the 2019 version).
We focused on very small young remnants and those containing x-ray point sources
or small pulsar wind nebulae.
We selected only those \acp{SNR} with age and distance estimates resulting in
$h_0^\mathrm{age}$ large enough to be detectable within our computing budget
(see below).
In addition to the Green \acp{SNR} we included the candidate
SNR~G354.4+0.0~\cite{Roy2013} as in Ref.~\cite{O1FifteenSNRs}, although a recent
multi-instrument comparison~\cite{Hurley-Walker2019} argues that it is probably
an HII region.
As in Ref.~\cite{O1FifteenSNRs}, we included SNR~G1.9+0.3 although it is
probably Type~Ia.
On the scale of our analysis, including two targets which might not contain
neutron stars added relatively little to the computational cost.

This process yielded the same 15 \acp{SNR} studied in the \ac{O1}
search~\cite{O1FifteenSNRs}.  We did not perform searches on
G111.7\textminus2.1, G266.2\textminus1.2, and G347.3\textminus0.5 from that
target list since they had already been searched~\cite{Ming2019} with
greater sensitivity than we could achieve with our more limited computational
resources.  The targets
for our searches are summarized in Table~\ref{t:table1}, along with sources of
their key astrophysical parameters.  Brief
descriptions and more details on the provenance of parameters
are given in~\cite{O1FifteenSNRs}.  
For four targets we ran ``wide'' and ``deep'' searches based on optimistic and
pessimistic estimates of age and distance from the literature, and thus we had
16 searches for 12 \acp{SNR}.
(Although the wide and deep searches cover the same frequencies, they cover
ranges of spin-down parameters that usually have little to no overlap.)
For G15.9+0.2 and G330.2+1.0 the deep searches were new---the \ac{O1} searches
could meet the sensitivity goal only for the optimistic estimates, but \ac{O2}
data allowed us to meet it even for pessimistic estimates.

Consistency checks on the parameters used were much easier than in
Ref.~\cite{O1FifteenSNRs}.
Here we used $f_{\max}$ of 150\,Hz, lower than in previous searches of this
type.
Hence errors due to neglect of higher frequency derivatives and other
approximations were reduced by a factor of a few to orders of magnitude over
previous searches, and were completely negligible.

\subsection{Computations}

Our searches used code descended from the pipeline used in some LIGO
searches~\cite{S5CasA, S6NineSNRs, O1FifteenSNRs} whose workhorse is the
$\mathcal{F}$-statistic as implemented in the \texttt{S6SNRSearch} tag of the
LALSuite software package~\cite{LALSuite}.
Search pipeline improvements mainly consisted of ``internal'' issues such as
better use of disk space, better error tracking, and improved interaction with
the batch job queuing system to reduce human workload.
Some significant bugs and issues were also addressed, as described below.

All searches ran on the Broadwell Xeon processors of the Quanah computing
cluster at Texas Tech.
Integration spans were adjusted by hand so that each search took approximately
$10^5$ core-hours, split into $10^4$ batch jobs.
Due to the frequency band used for the searches, which avoided the worst
spectrally disturbed bands, the total search output used less than one terabyte
of disk space.

\begin{figure*}
  \begin{center}
    \includegraphics[width=0.48\textwidth]{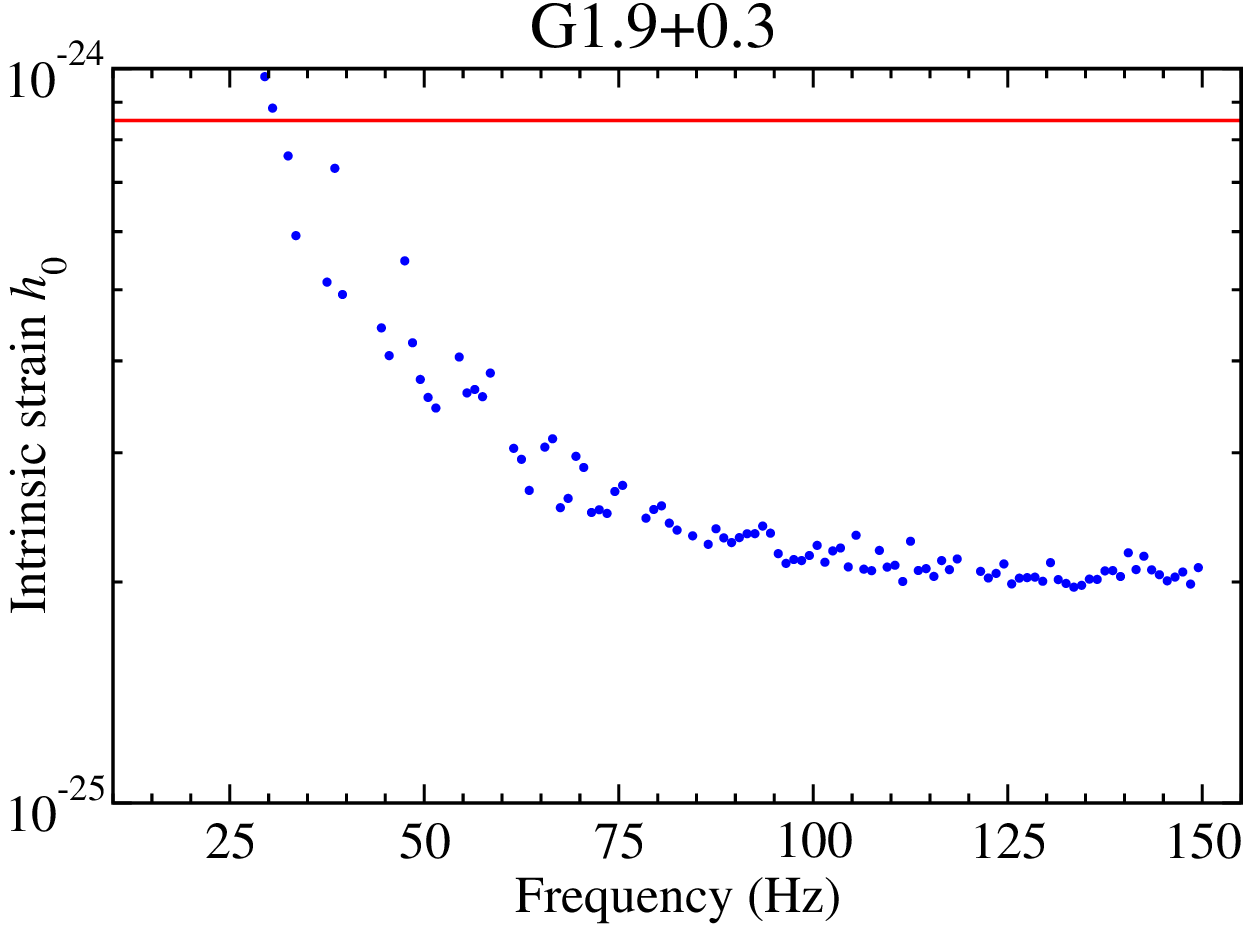}
    \hspace{0.25cm}
    \includegraphics[width=0.48\textwidth]{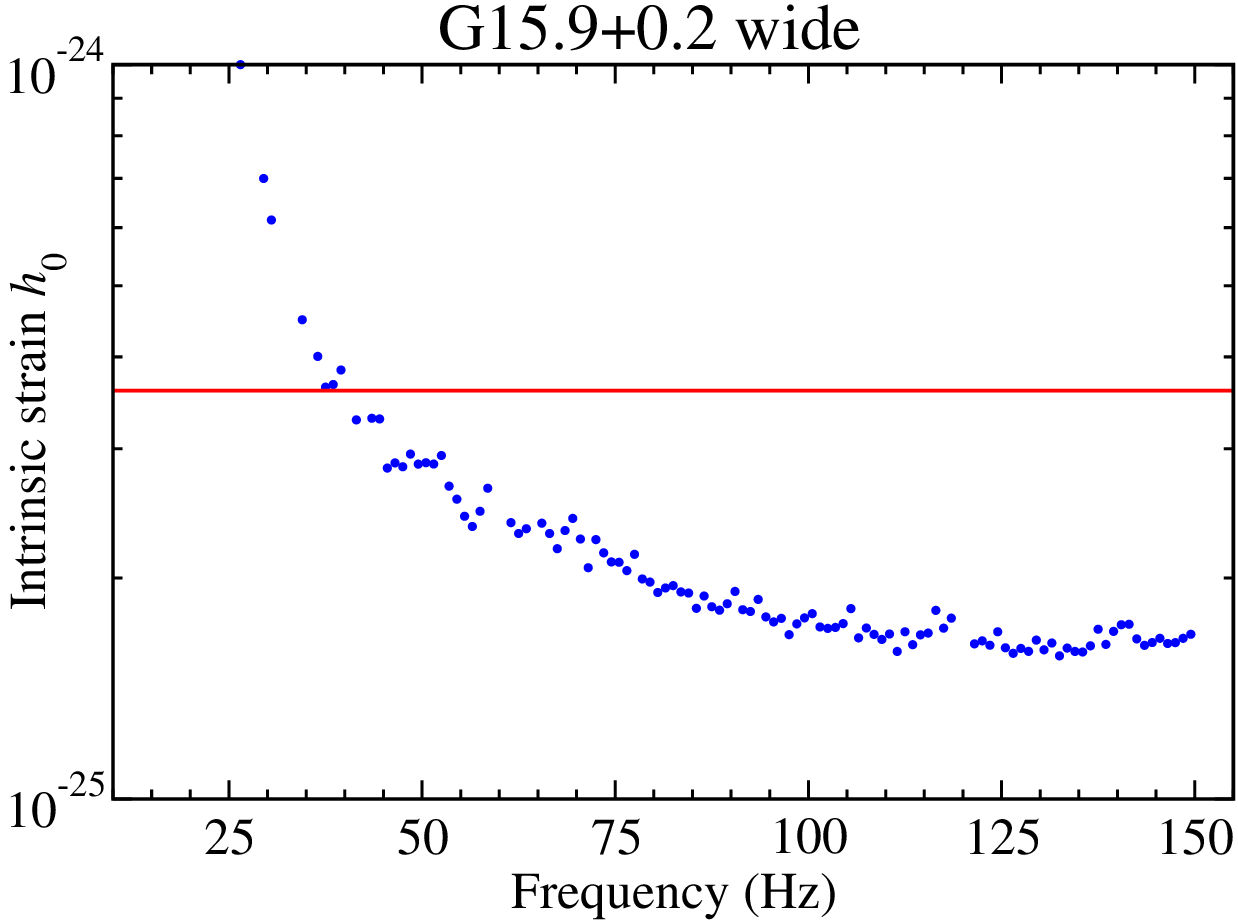}
    
    \vspace{10pt}
    \includegraphics[width=0.48\textwidth]{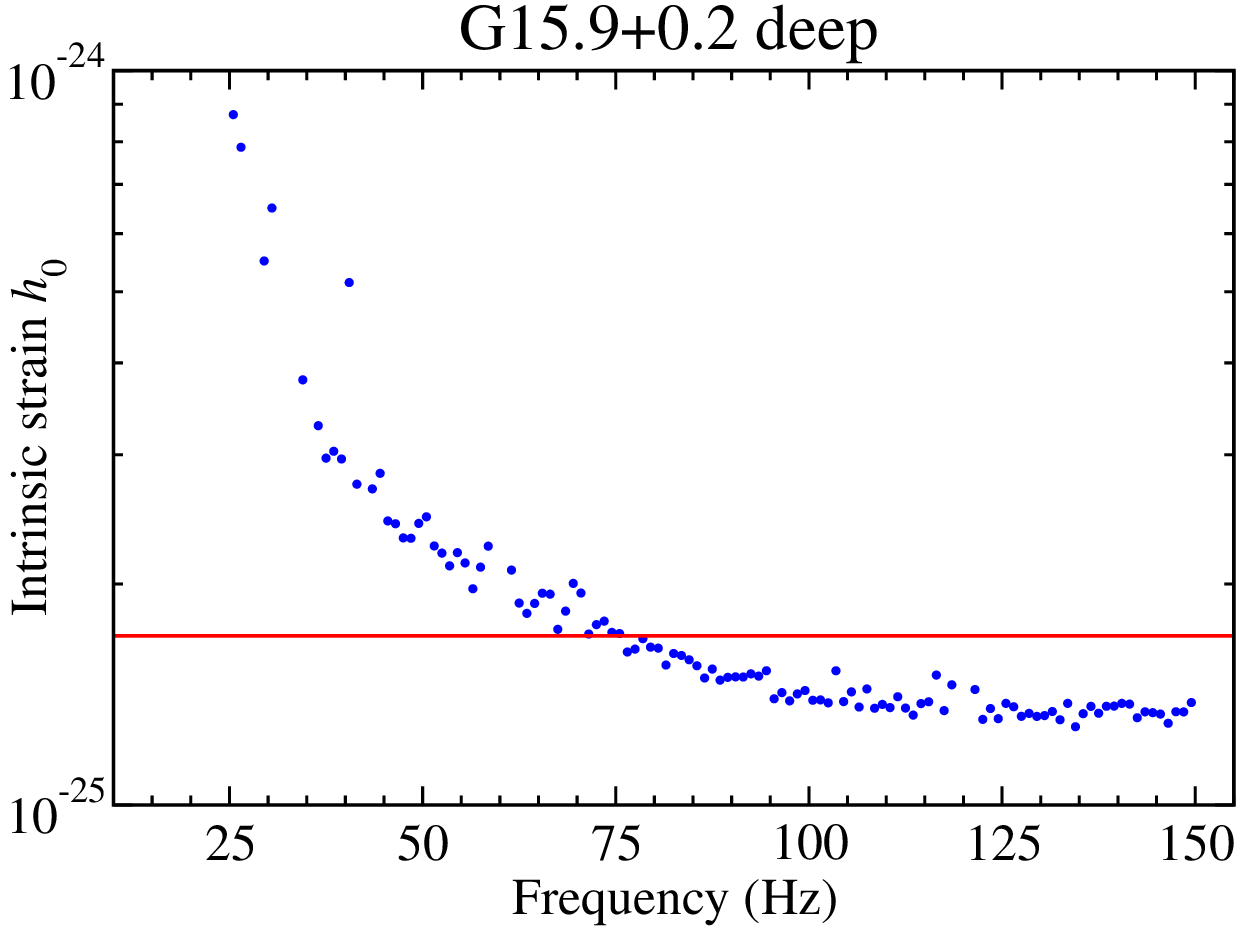}
    \hspace{0.25cm}
    \includegraphics[width=0.48\textwidth]{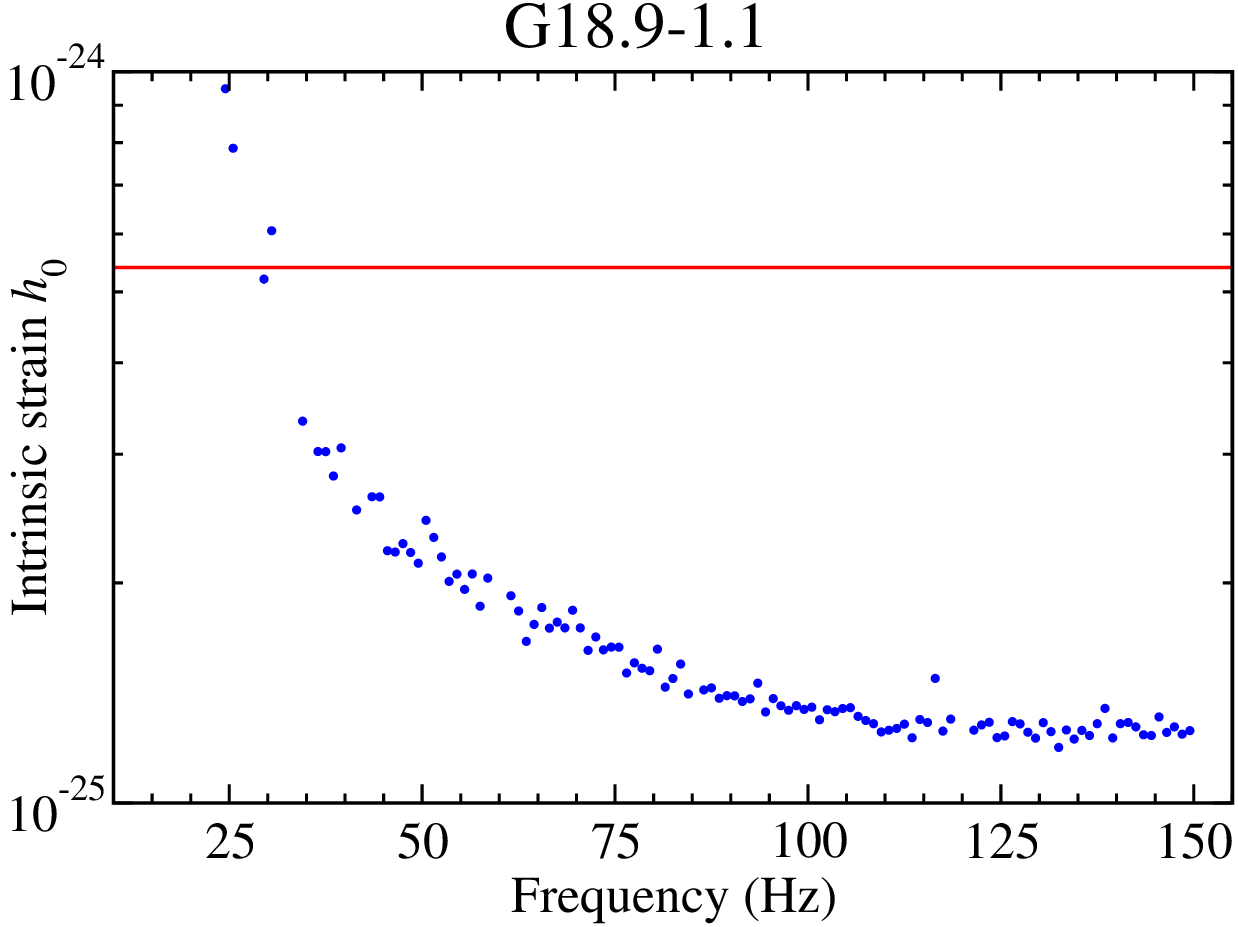}
  \end{center}
  \caption{\label{f:Fig1} Points represent the direct observational
    90\% confidence upper limits on the intrinsic strain $h_0$ as a
    function of frequency in 1~Hz bands for four searches.  The (red)
    horizontal line indicates the indirect limit
    $h_0^\mathrm{age}$ from
    energy conservation.  All figures trace a slightly distorted
    version of the noise curve.}
\end{figure*}

\begin{figure*}
  \begin{center}
    \includegraphics[width=0.48\textwidth]{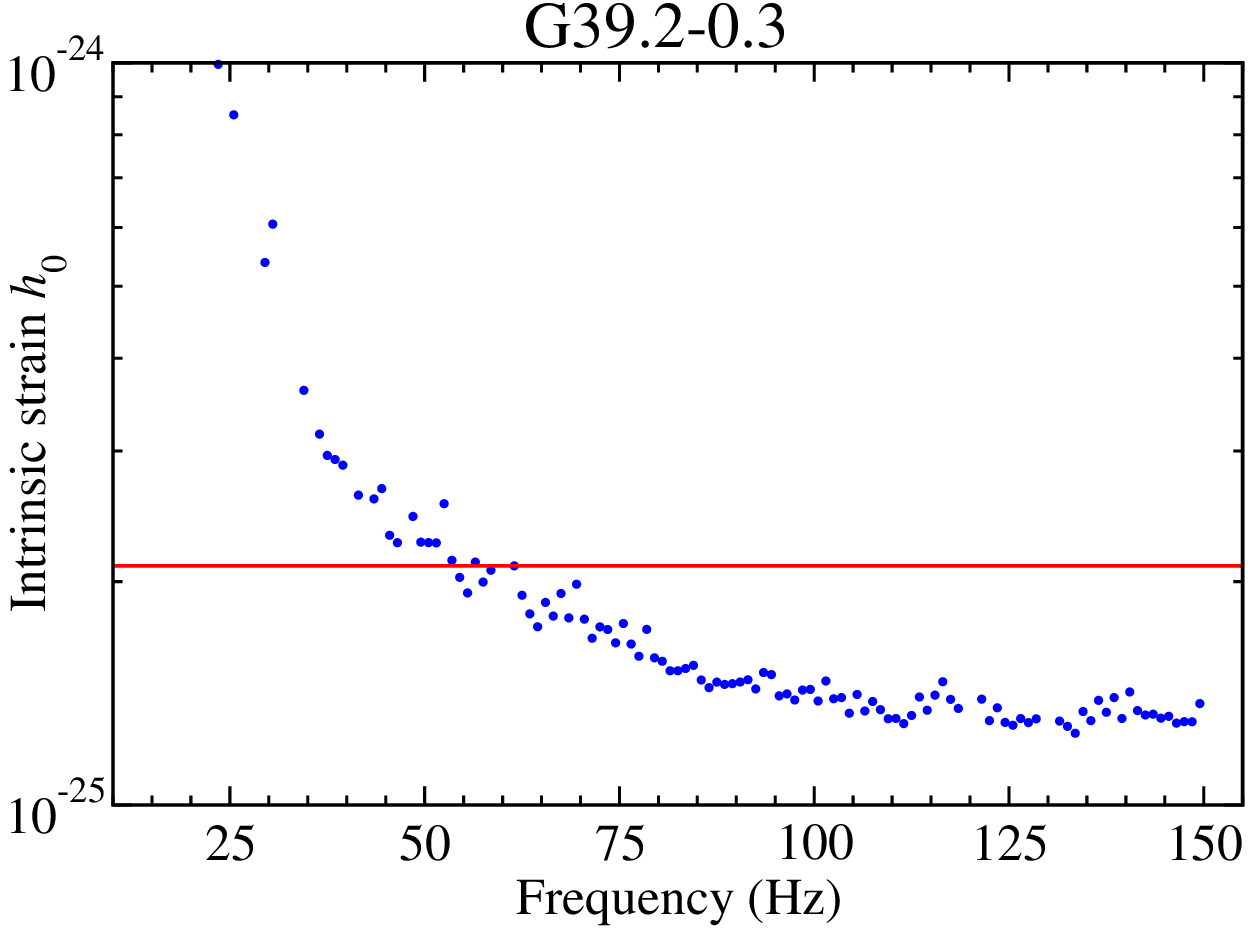}
    \hspace{0.25cm}
    \includegraphics[width=0.48\textwidth]{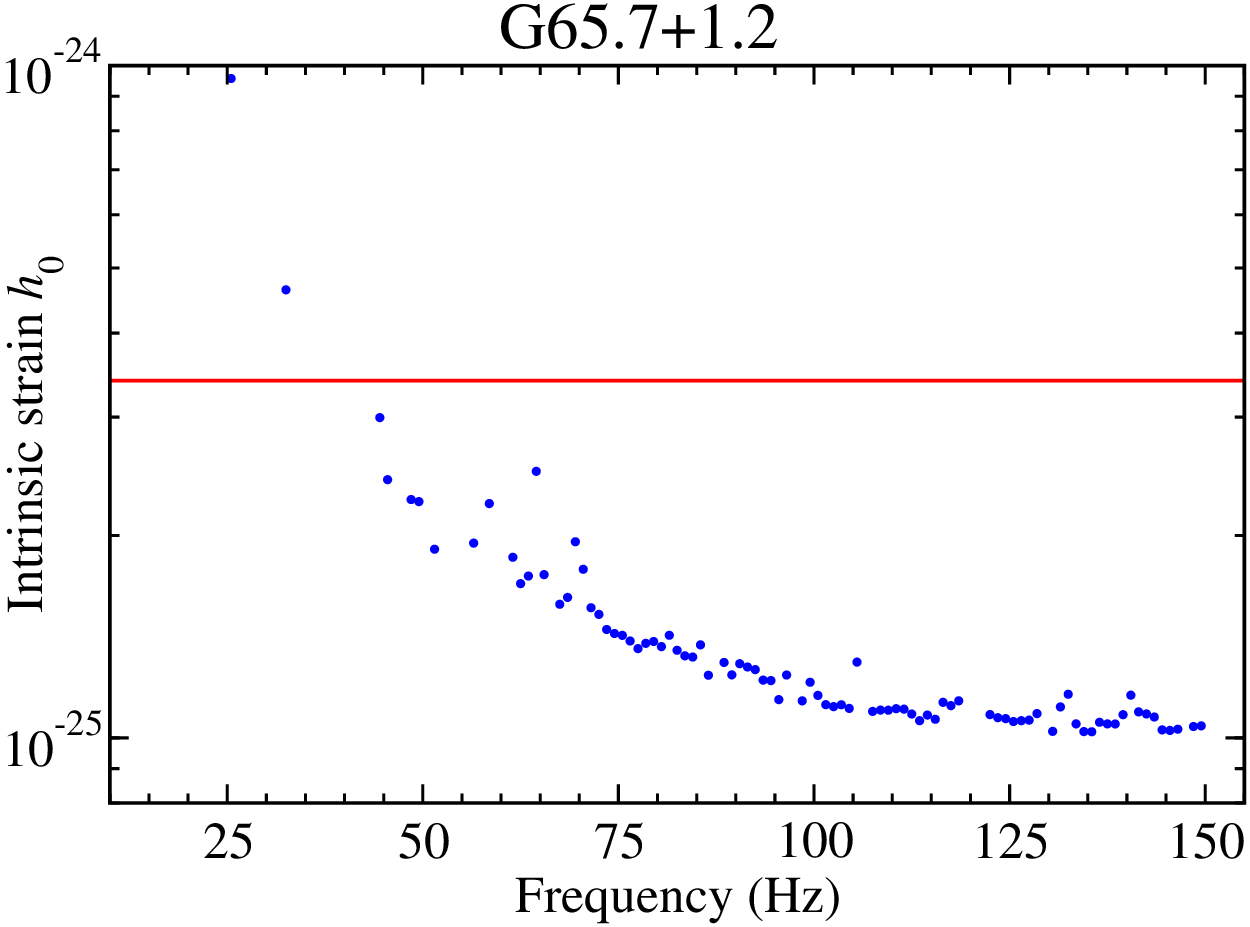}
    
    \vspace{10pt}
    \includegraphics[width=0.48\textwidth]{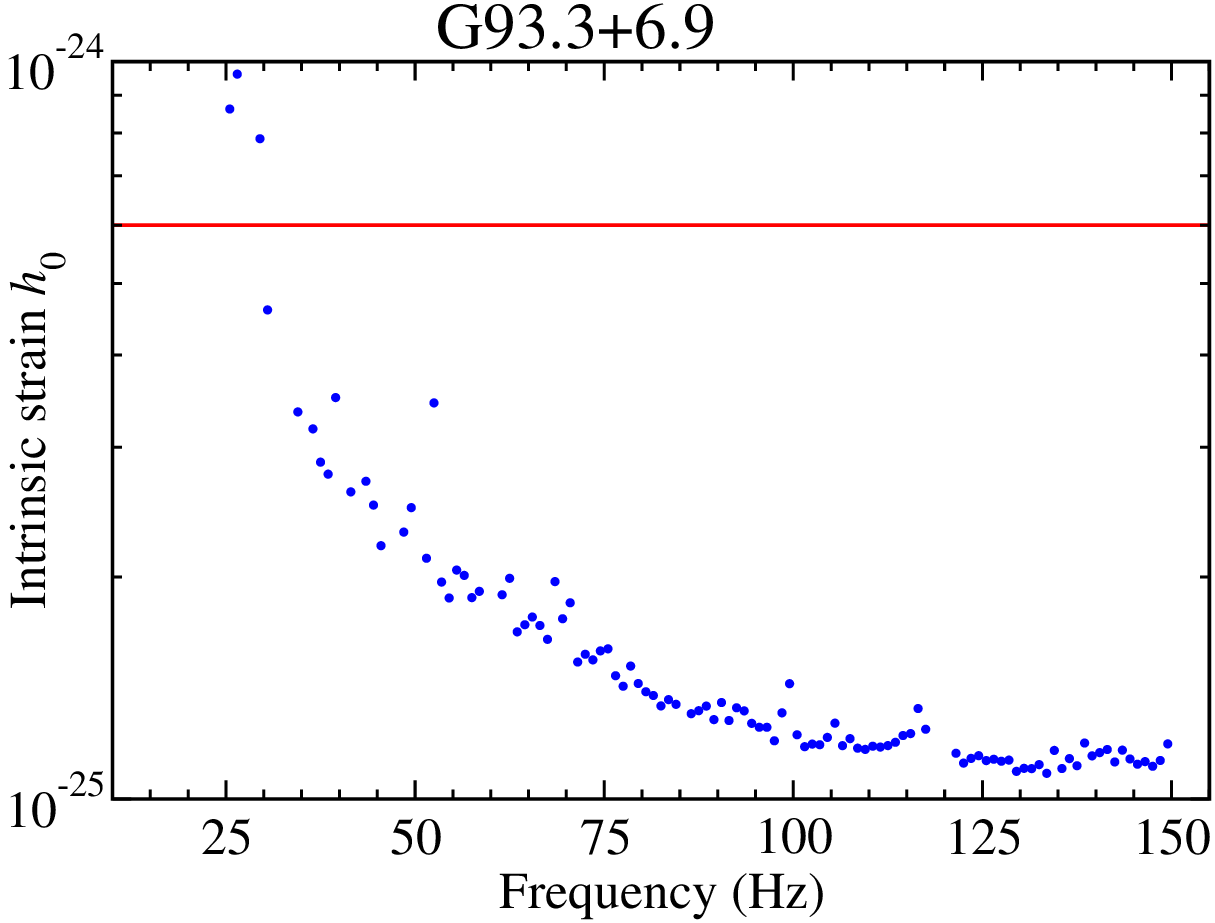}
    \hspace{0.25cm}
    \includegraphics[width=0.48\textwidth]{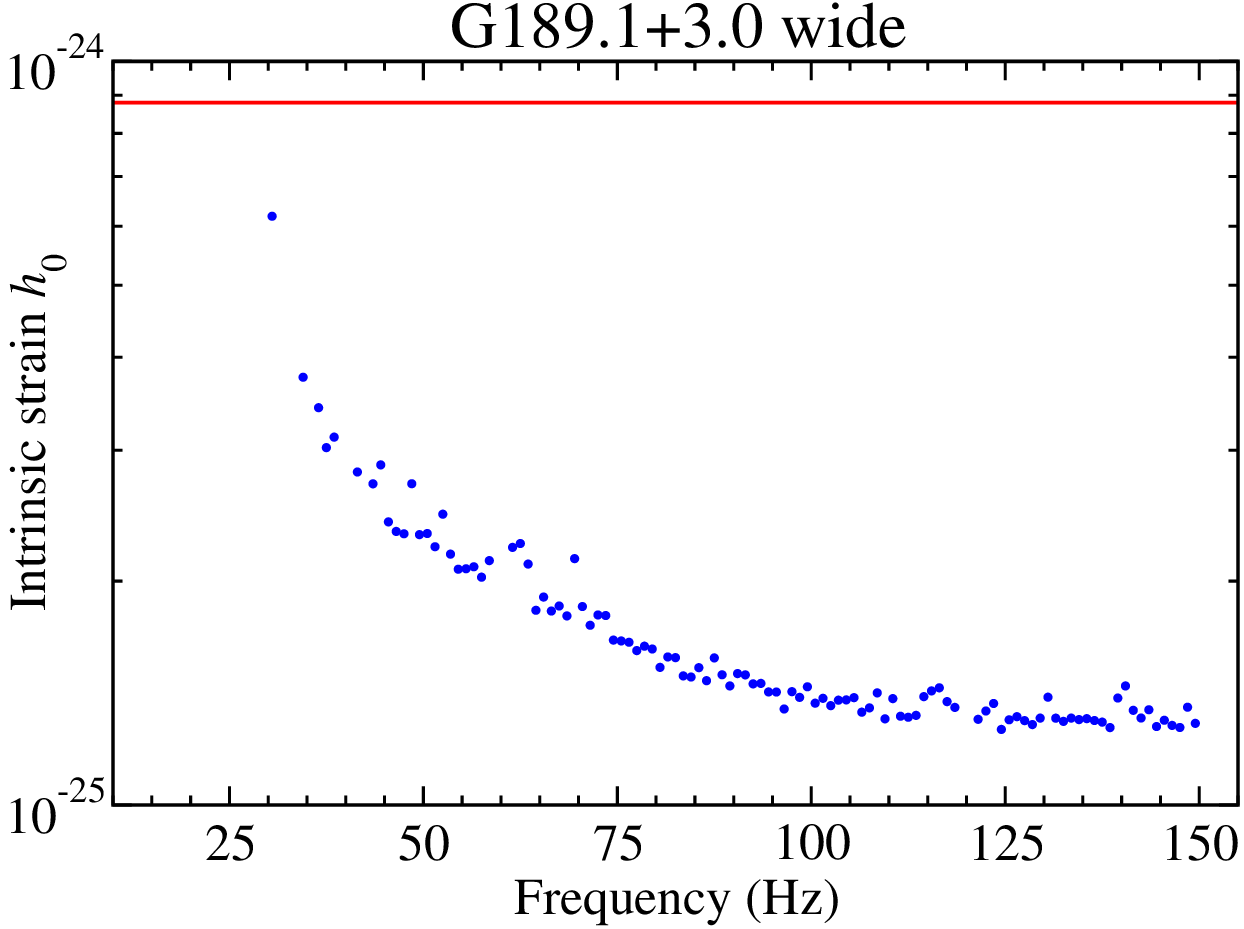}
  \end{center}
  \caption{\label{f:Fig2} Same as Fig.~\ref{f:Fig1} for four
    additional searches.}
\end{figure*}

\subsection{Post-processing}

As in the \ac{O1} search~\cite{O1FifteenSNRs}, post-processing of search results
started with the ``Fscan veto'' and interferometer consistency veto.
The former uses a normalized spectrogram to check for spectral lines and
nonstationary noise.
The latter checks that the two-interferometer $\mathcal{F}$-statistic is greater
than the value of either single-interferometer $\mathcal{F}$-statistic; failure
of this condition strongly indicates a spectral line.

We found and fixed several bugs in the post-processing part of the pipeline.
Their total effect on previous searches was negligible (the false dismissal rate
of Ref.~\cite{O1FifteenSNRs} was wrong by a few times 0.01\%).
However the effect on previous upper limits was more substantial, as described
in the next Section.

The \ac{O1} pipeline~\cite{O1FifteenSNRs} corrected a bug in earlier
versions~\cite{S5CasA, S6NineSNRs} whereby the Doppler shift due to
the Earth's orbital motion was omitted when applying detector-frame
vetoes to candidate signals whose frequency is recorded in the solar
system barycenter frame.  However, we found that in the process the
\ac{O1} pipeline introduced a bug in which $\dot f$ and $\ddot f$ were
ignored when computing the frequency bands affected by the Fscan veto
and removal of known lines. Although we did not remove known lines, we
found that this bug fix reduced the number of candidate signals.  This
also means that the searches in Ref.~\cite{O1FifteenSNRs} spuriously
vetoed a fraction of the frequency band on the order of
$T_\mathrm{span}/a$ for each search, of order a few times $10^{-4}$
for the worst case (SNR~G1.9+0.3), increasing the false dismissal rate
by about that (negligible) amount..

When the $\dot f$-$\ddot f$ bug was fixed, it significantly increased the total
frequency band vetoed in each search.
As before, the veto criterion was very strict, including all templates whose
detector-frame frequency ever came within eight \ac{SFT} bins (almost 5\,mHz) of
an Fscan with sufficiently high power.
(Eight bins was the width of the Dirichlet kernel used in computing the
$\mathcal{F}$-statistic.)
However, in the interest of setting upper limits on most frequency bands, we
raised the power threshold (loosened the veto) from seven standard deviations to
twenty.
This brought the total vetoed band back down comparable to what it was in
previous analyses such as Ref.~\cite{O1FifteenSNRs}.
As we shall see, this helped us set upper limits broadly without letting through
an onerous number of candidate signals for manual inspection.

Unlike Ref.~\cite{O1FifteenSNRs} we did not veto using the list of known
instrumental lines~\cite{Covas2018}.
With the $\dot f$-$\ddot f$ bug fixed, the total frequency band vetoed would
have significantly reduced the number of upper limits we could set at high
confidence.
Also, we found that the search and bug-fixed Fscan veto performed quite well on
most lines.

After these automated data-based vetoes were applied, the pipeline produced 21
search jobs whose loudest nonvetoed $\mathcal{F}$-statistic exceeded the 95\%
confidence threshold for gaussian noise.
We inspected all these candidates using the criteria from
Ref.~\cite{O1FifteenSNRs}, essentially looking at the frequency spectrum of each
candidate and the candidate's effect on the histogram of $\mathcal{F}$-statistic
values.

No candidate survived visual inspection---all were much too broad-band compared
to hardware-injected pulsar signals and had distorted histograms.
Although we did not use the known lines as \textit{a priori} vetoes, we checked
\textit{a posteriori} and found that most candidates were related to harmonics
of 60\,Hz or 0.5\,Hz or to hardware-injected pulsar six, which was found
(slightly Doppler shifted and broadened) in multiple searches at different sky
locations.
The \ac{O2} injected pulsar parameters are listed in Ref.~\cite{O2CWAllSky}.
Although it was loud, for many searches injected pulsar six was not loud enough
to trigger the Fscan veto.

\begin{figure*}
  \begin{center}
    \includegraphics[width=0.48\textwidth]{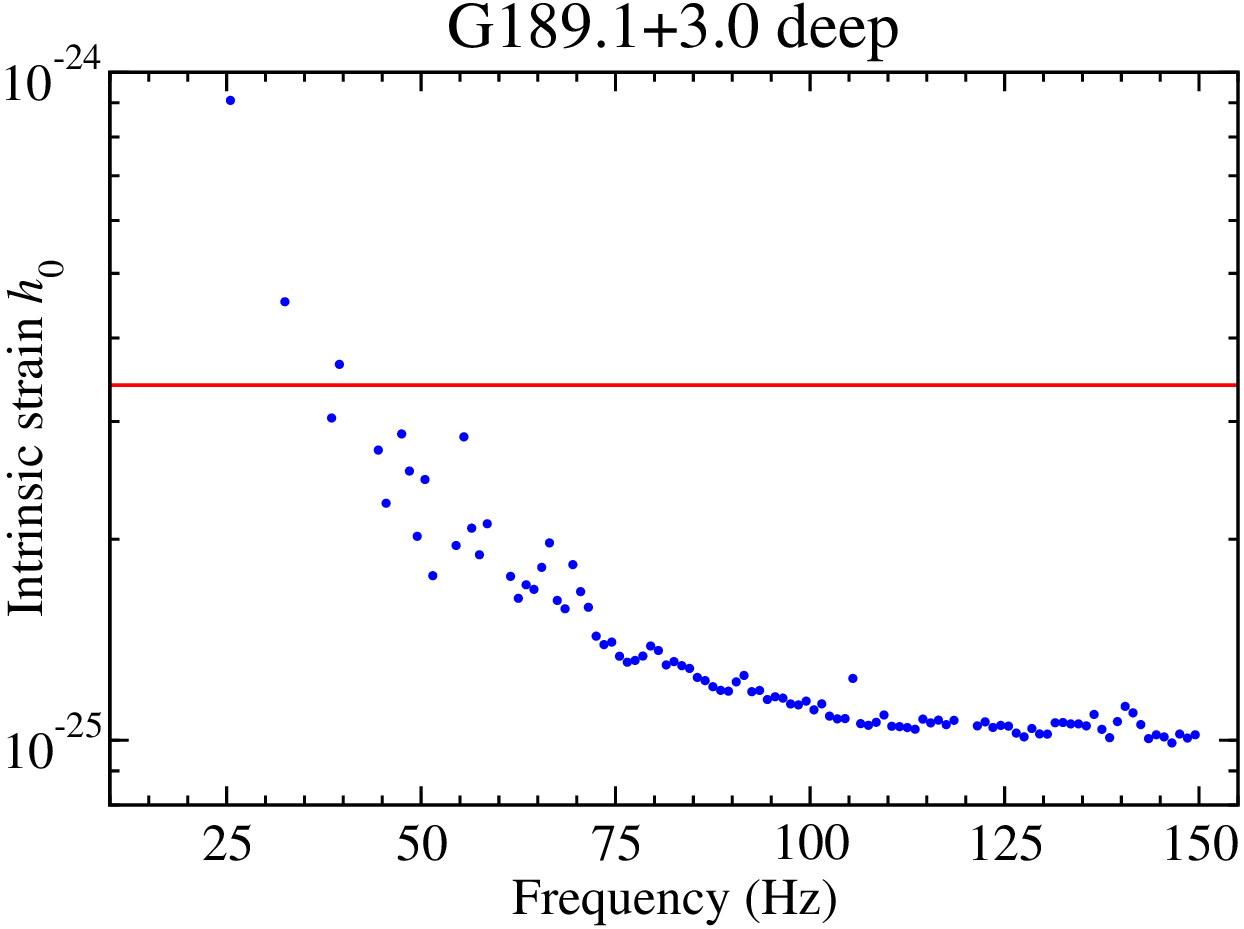}
    \hspace{0.25cm}
    \includegraphics[width=0.48\textwidth]{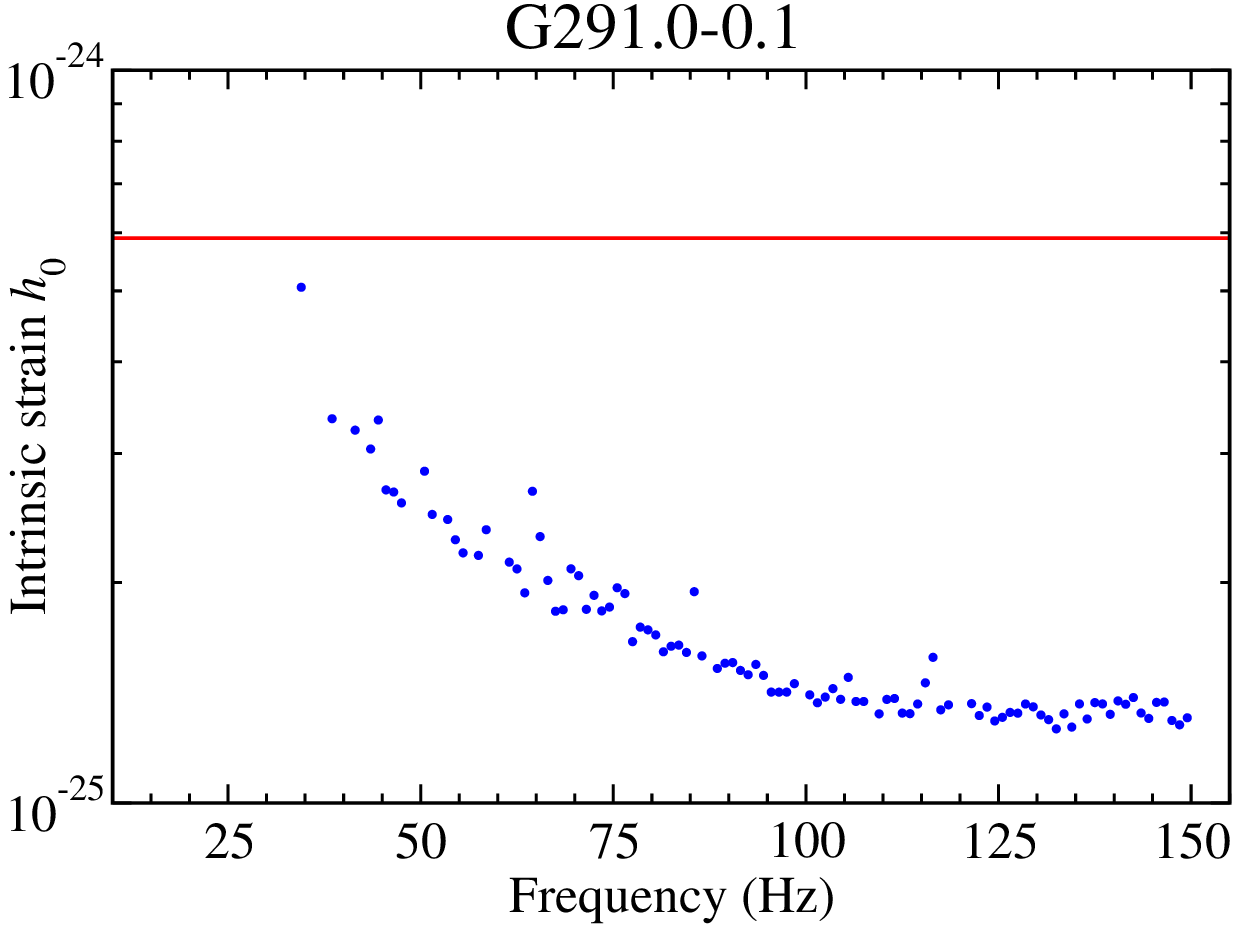}
    
    \vspace{10pt}
    \includegraphics[width=0.48\textwidth]{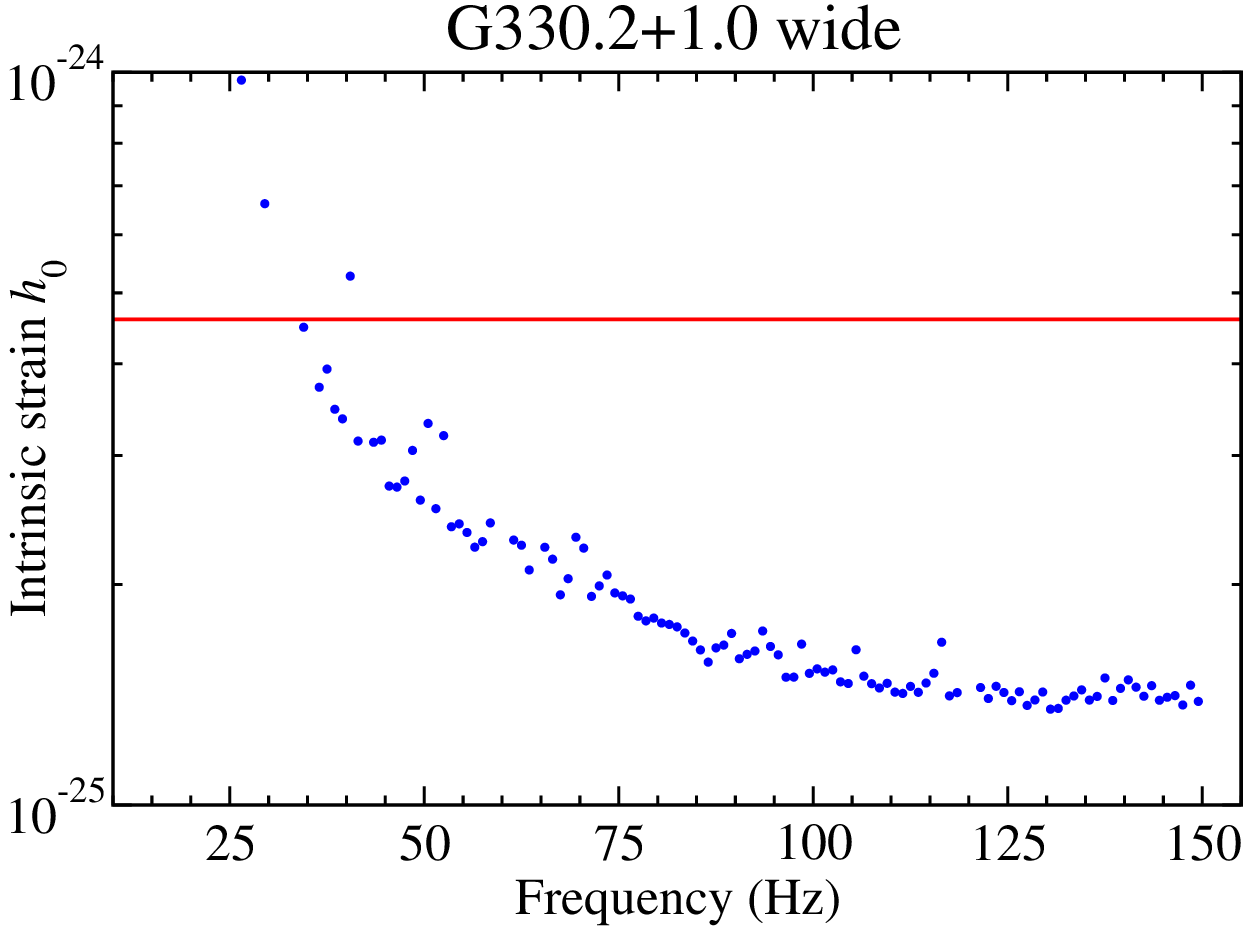}
    \hspace{0.25cm}
    \includegraphics[width=0.48\textwidth]{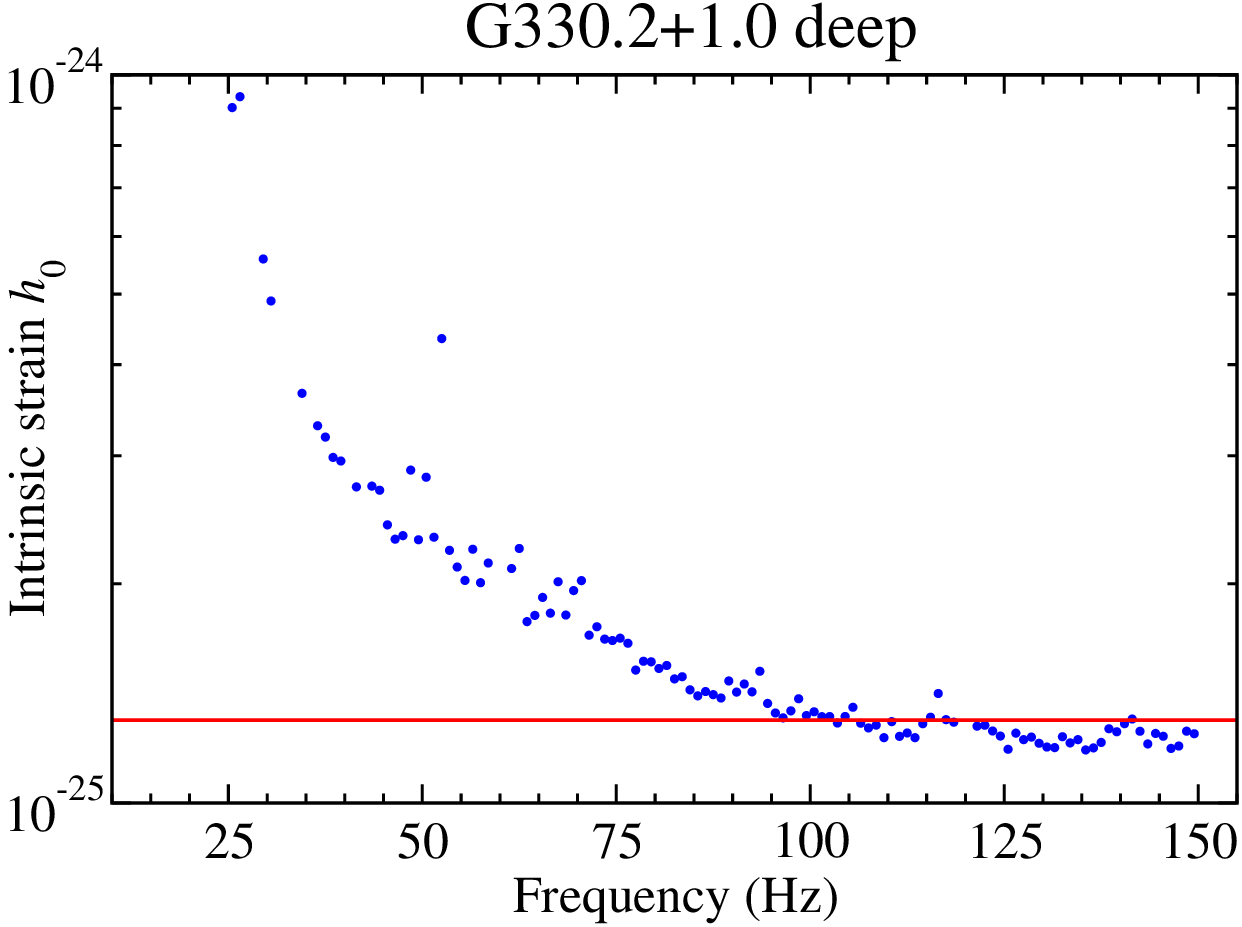}
  \end{center}
  \caption{\label{f:Fig3} Same as Fig.~\ref{f:Fig1} for four
    additional searches.}
\end{figure*}

\section{Upper Limits}

Having detected no signals, we placed upper limits on $h_0$ in 1\,Hz bands using
a procedure similar to Ref.~\cite{O1FifteenSNRs}.
That, is we estimated the $h_0$ that would be detected in each band (with the
$\mathcal{F}$-statistic louder than the loudest actually recorded in that band)
with a certain probability if the other signal parameters were varied randomly.
This estimate used semi-analytical approximations to the $\mathcal{F}$-statistic
probability distribution integrals and was spot-checked using one thousand
software-injected signals per upper limit band.

Unlike~\cite{O1FifteenSNRs}, which set upper limits at 95\% confidence, we
reduced the confidence level to 90\% (10\% false dismissal).
This was necessary to reduce the number of bands unsuitable for an upper limit.
Upper limit bands were deemed unsuitable and dropped if more than 10\% of the
band was vetoed.
We also dropped bands immediately adjoining 60\,Hz and 120\,Hz, the fundamental
and first overtone of the electrical power mains.
By spot checking the upper limit injections we found that, all else being equal,
changing the confidence from 95\% to 90\% reduced the $h_0$ upper limits by
5--8\%.
This difference is less than the calibration errors and negligible for the
purposes of comparing to previous work.

Related to this, we found a bug in the \ac{O1} code whereby known line vetoes
were not included in the total band veto.
Since there were many known lines, this and the $\dot{f}$-$\ddot{f}$ bug meant
that the vetoed band totals in Ref.~\cite{O1FifteenSNRs} were often greatly
underestimated.
Strictly speaking, perhaps half of the 95\% upper limit points should have been
dropped.
Or they should have used 90\% confidence as we do here, which would have changed
$h_0$ by a few percent.

\begin{figure*}
  \begin{center}
    \includegraphics[width=0.48\textwidth]{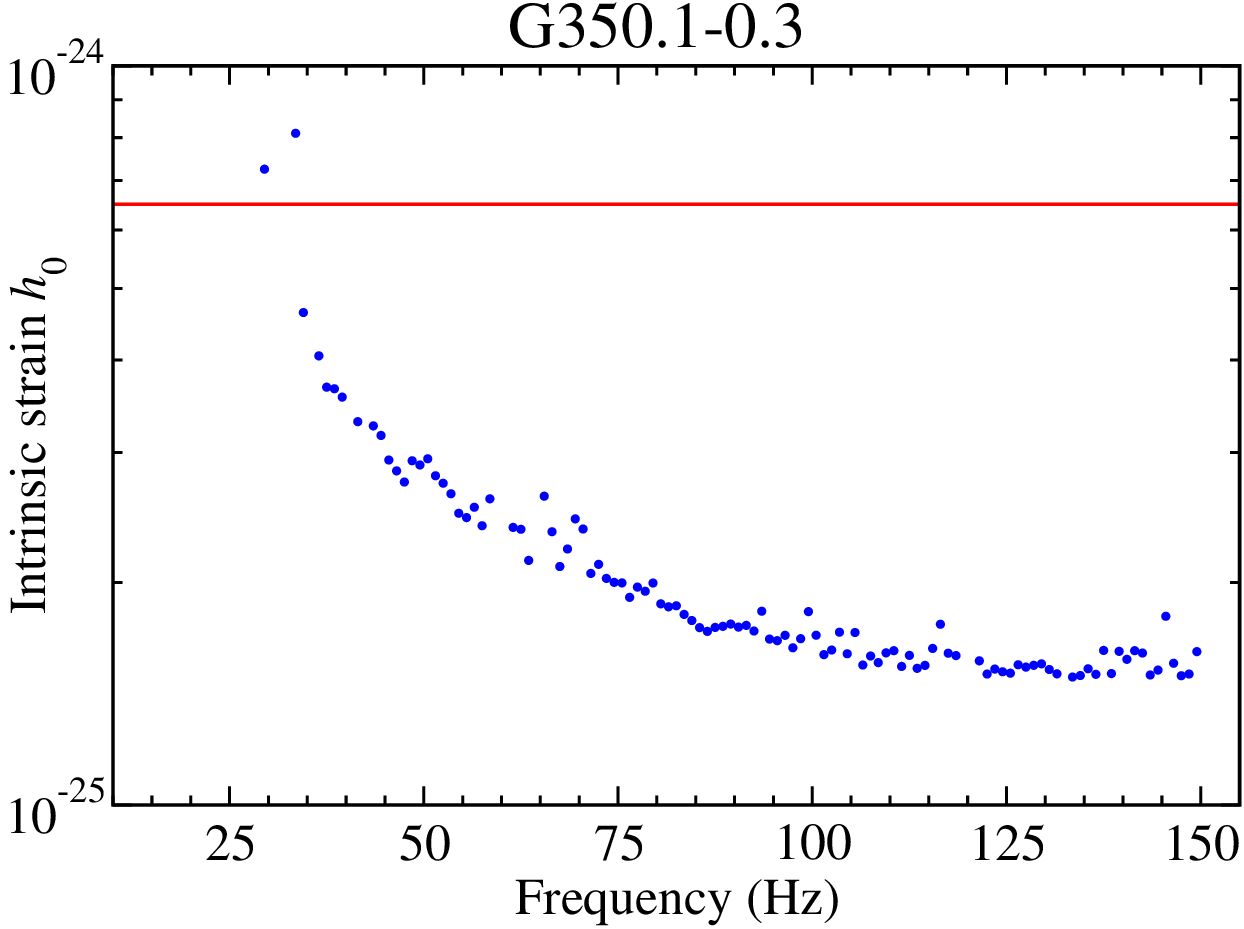}
    \hspace{0.25cm}
    \includegraphics[width=0.48\textwidth]{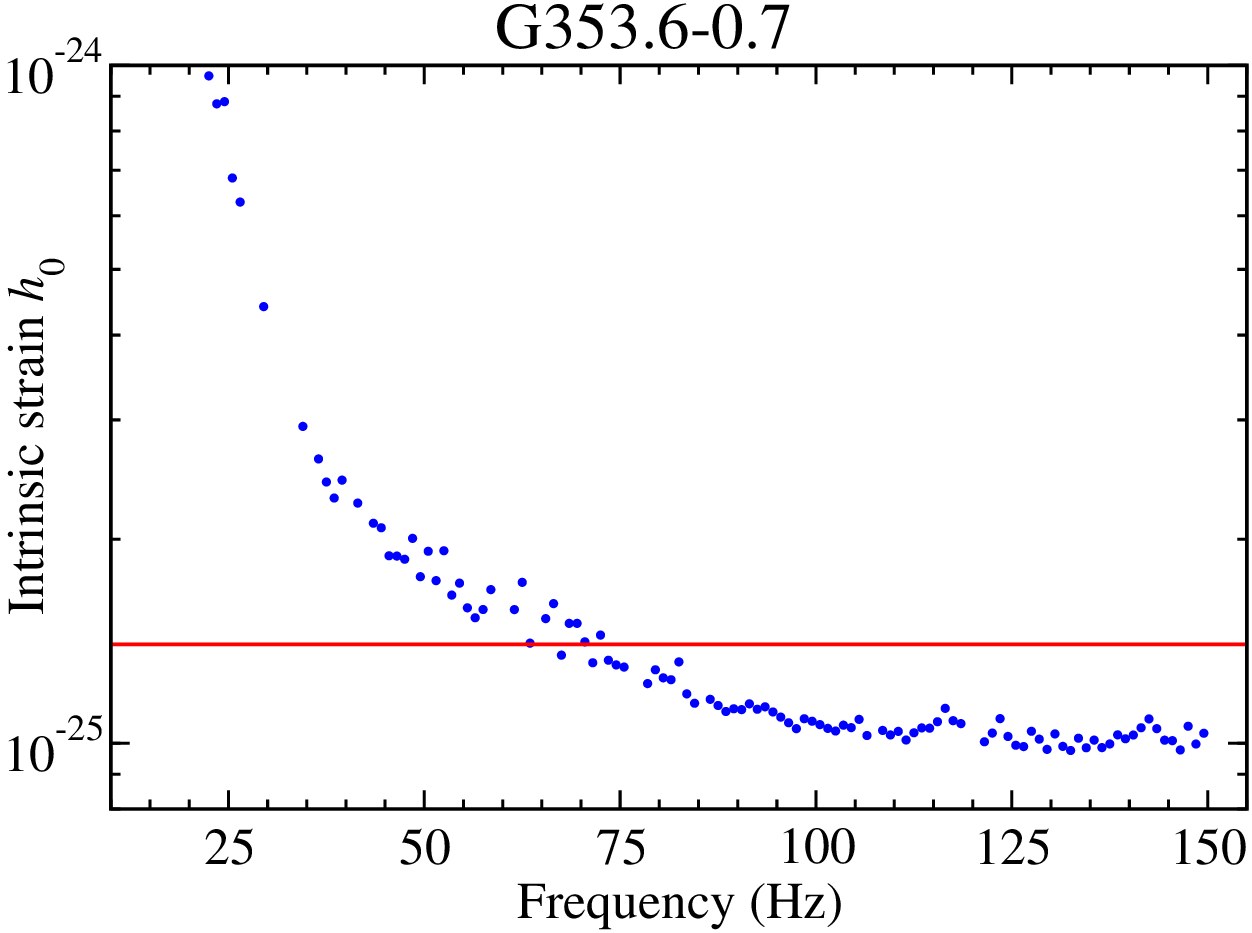}
    
    \vspace{10pt}
    \includegraphics[width=0.48\textwidth]{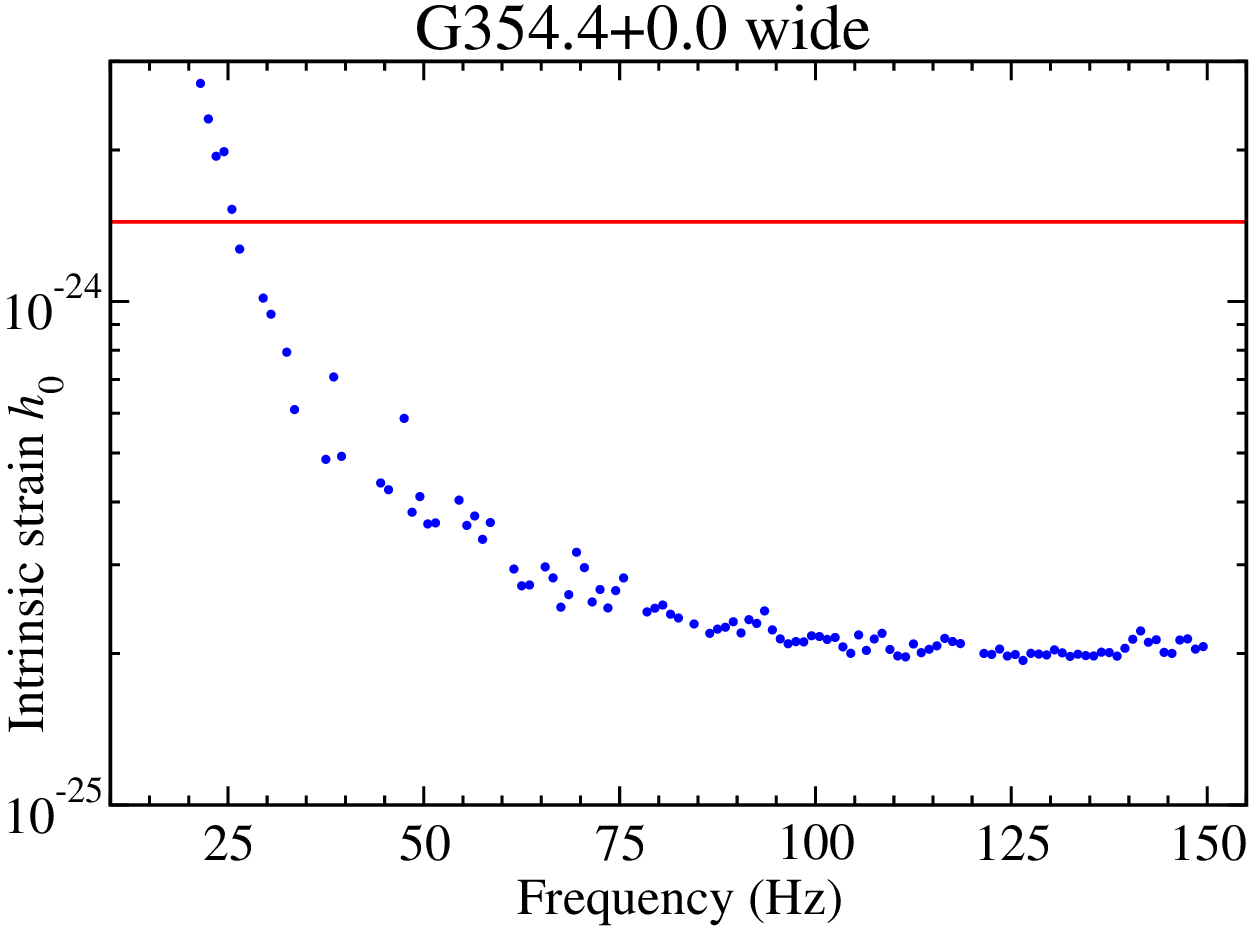}
    \hspace{0.25cm}
    \includegraphics[width=0.48\textwidth]{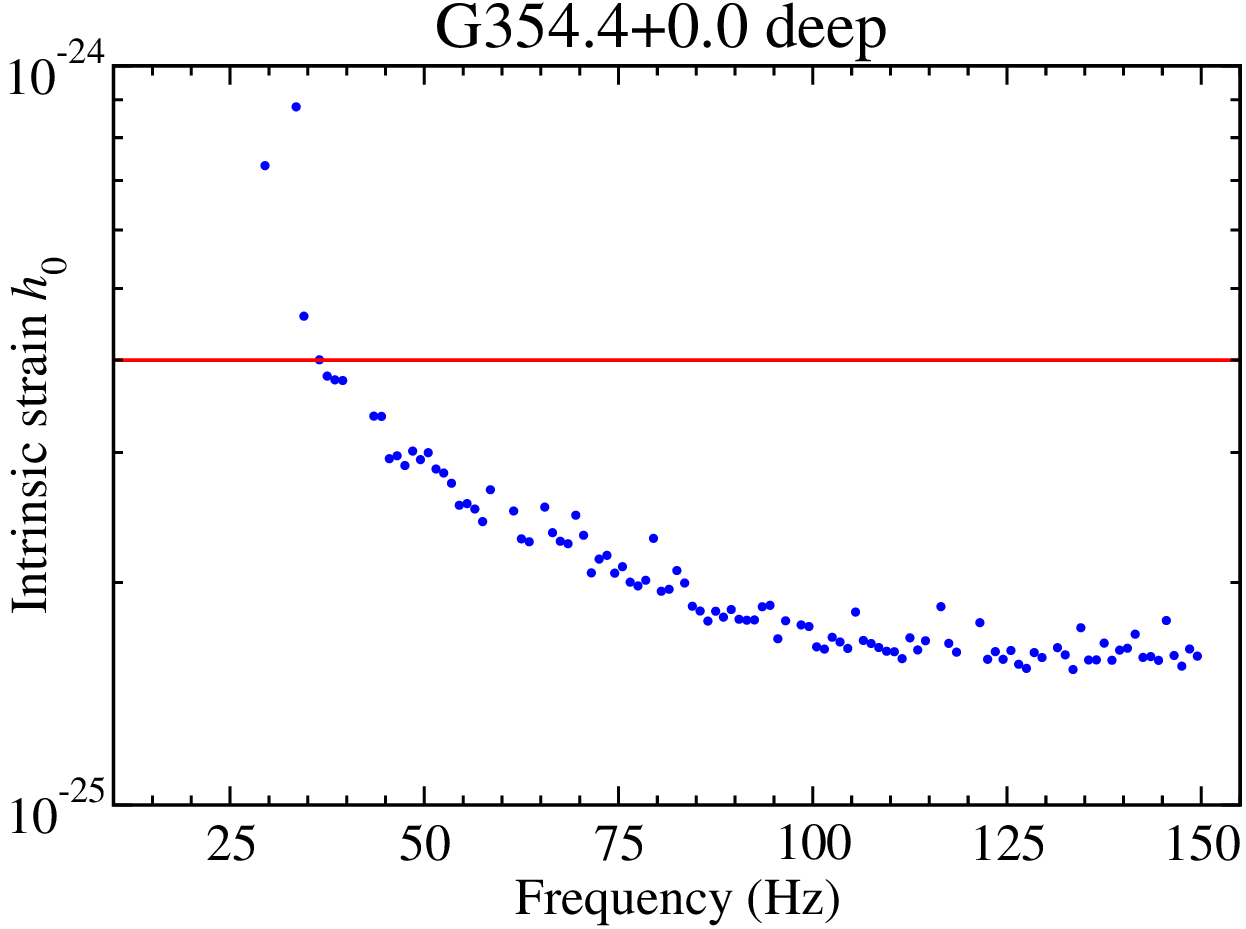}
  \end{center}
  \caption{\label{f:Fig4} Same as Fig.~\ref{f:Fig1} for four
    additional searches. }
\end{figure*}

The upper limits on $h_0$ which survived the veto check are plotted as a
function of frequency in Figs.~\ref{f:Fig1}--\ref{f:Fig4}.
Generally older targets produced better upper limits because longer integration
times were possible for the fixed computational cost per target.
The data files, including points not visible on the plots, are included in the
supplemental material to this article~\cite{EPAPS}.
In terms of the ``sensitivity depth'' defined in Ref.~\cite{Dreissigacker2018},
these searches ranged from about 45\,Hz$^{-1/2}$ for young \acp{SNR} to
70\,Hz$^{-1/2}$ for older \acp{SNR}.

Upper limits on $h_0$ can be converted to upper limits on neutron star
ellipticity $\epsilon$ using e.g.~\cite{Wette2008}
\begin{equation}
\epsilon \simeq 9.5\times10^{-5} \left( \frac{h_0} {1.2\times10^{-24}} \right)
\left( \frac{D} {\mbox{1 kpc}} \right) \left( \frac{\mbox{100 Hz}} {f}
\right)^2
\end{equation}
and to upper limits on a particular measure of $r$-mode amplitude
$\alpha$~\cite{Lindblom1998} using~\cite{Owen2010}
\begin{equation}
\alpha \simeq 0.28 \left( \frac{h_0} {10^{-24}} \right) \left( \frac{\mbox{100
Hz}} {f} \right)^3 \left( \frac{D} {\mbox{1 kpc}} \right).
\end{equation}
The numerical values are uncertain by a factor of roughly two or three due to
uncertainties in the unknown neutron star mass and equation of state.

\begin{figure*}
  \begin{center}
    \includegraphics[width=0.45\textwidth]{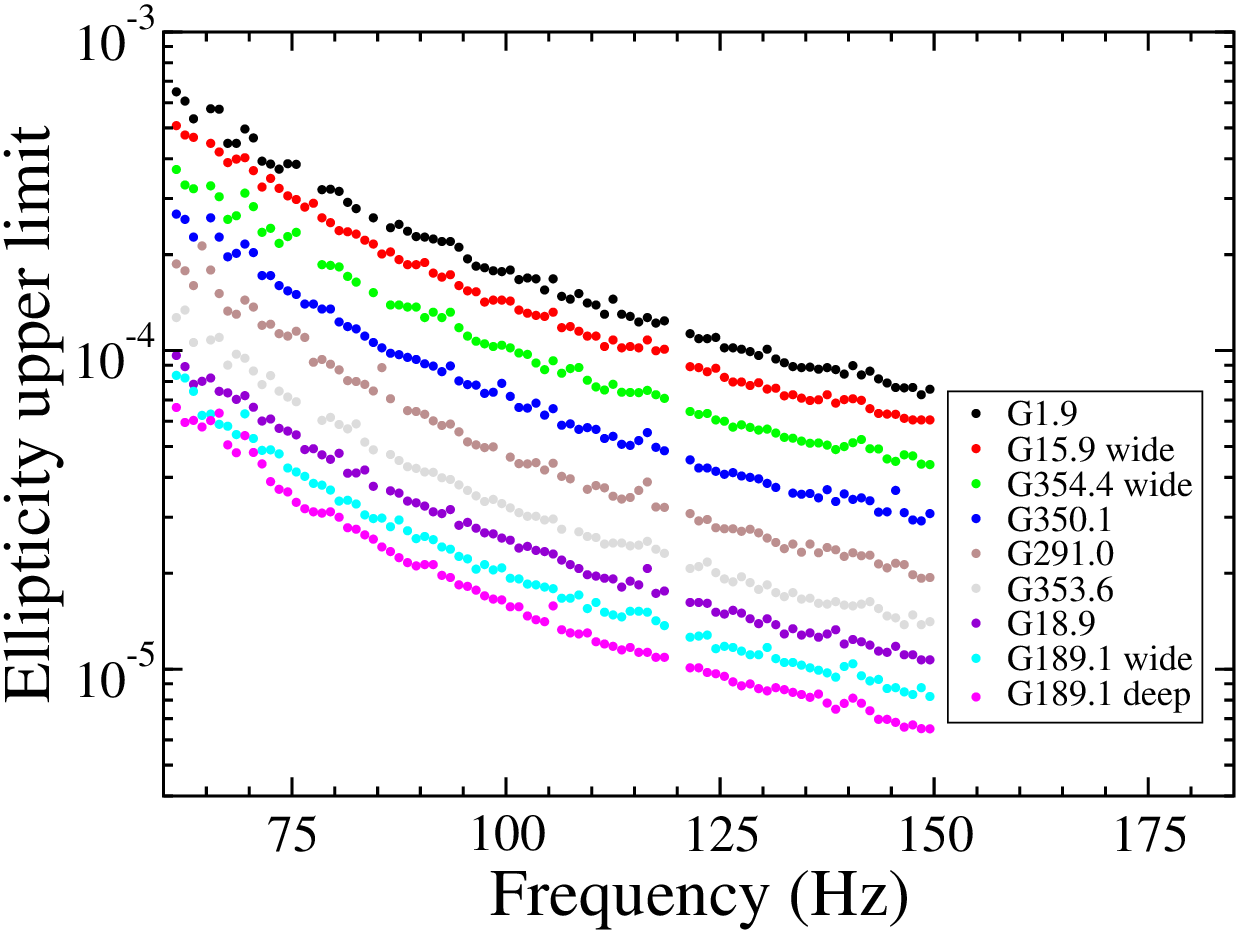}
    \hspace{0.25cm}
    \includegraphics[width=0.45\textwidth]{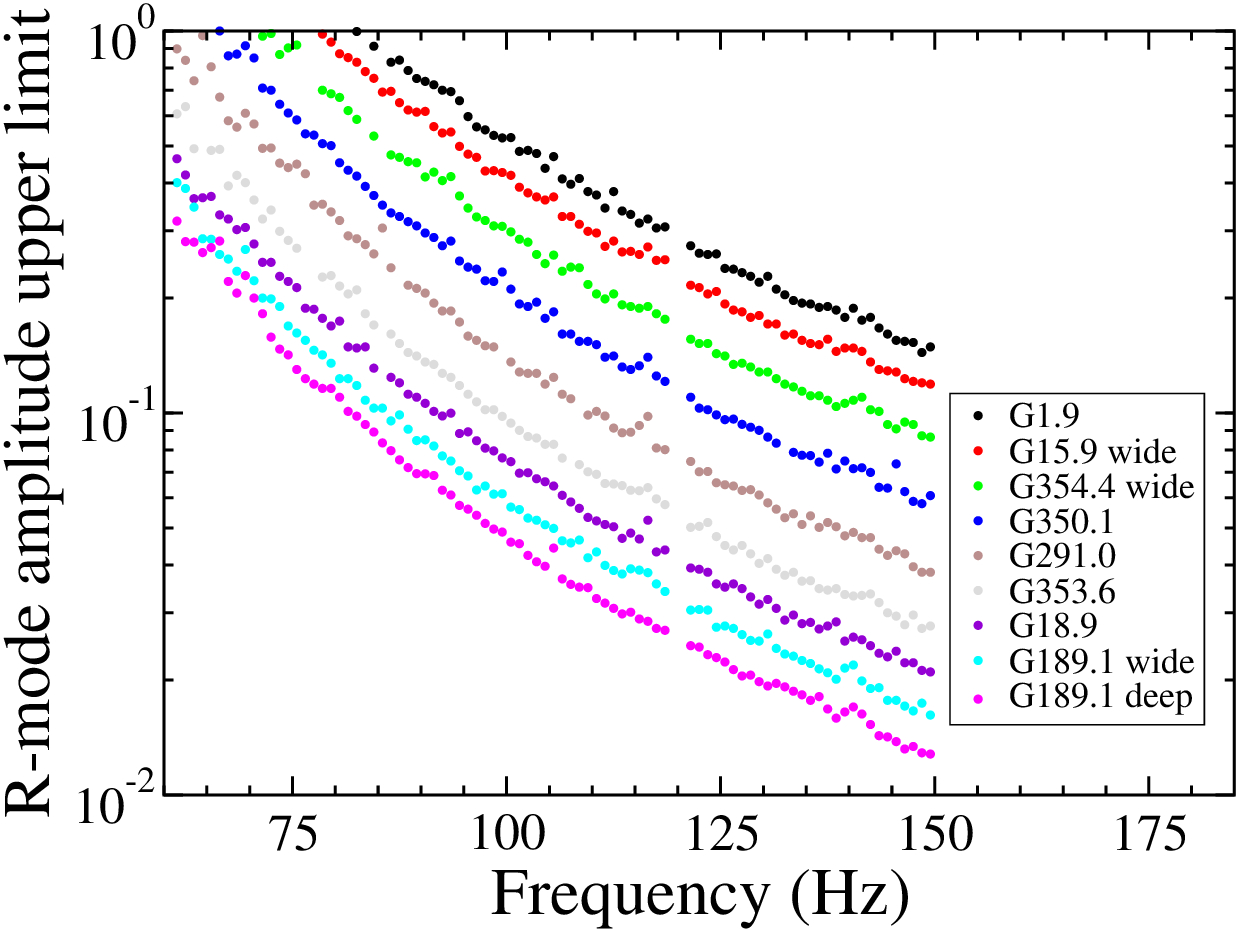}
  \end{center}
  \caption{\label{f:Fig6} Upper limits on fiducial neutron-star
    ellipticity (left panel) and $r$-mode amplitude (right panel) for a
    representative sample of \acp{SNR}.
    See the text for details.}
\end{figure*}

We plot upper limits on $\epsilon$ for a selection of searches representing the
range of these limits in the left panel of
Fig.~\ref{f:Fig6} and on $\alpha$ in the right panel.
The differences between curves are primarily due to to differences in the
distances to the sources.

\section{Discussion}

Although we detected no signals, we placed the best upper limits yet on \ac{GW}
amplitude from these twelve \acp{SNR}.
Our upper limits are as good (low) as $1.0\times10^{-25},$ and were generally
about a factor of two better than similar limits on the same \acp{SNR} using
\ac{O1} data from Ref.~\cite{O1FifteenSNRs}.
Our upper limits are also (for several targets) up to a factor of two better
than all-sky limits on \ac{O2} data from Ref.~\cite{O2CWAllSky}.
For SNR~G1.9+0.3 our limits were about the same as Ref.~\cite{O2CWAllSky} in our
frequency band, but we covered five times the range of $\dot f.$
Also, our searches included $\ddot f$ which is rare in the literature.
Our searches included two new parameter sets for two of the \acp{SNR}.
Part of the improved sensitivity over comparable \ac{O1}
searches~\cite{O1FifteenSNRs} was due to the reduced noise of \ac{O2} and part
was due to our longer searches at lower frequencies, which seem to be
characteristic of known young pulsars.
Because of our focus on lower frequencies, our upper limits on neutron-star
ellipticity and $r$-mode amplitude are less impressive than those from searches
which extended to higher frequencies~\cite{O1FifteenSNRs}.
Our limits on $r$-mode amplitude do not reach the $10^{-3}$ level expected by
the most detailed exploration of nonlinear saturation
mechanisms~\cite{Bondarescu2009}.
But our ellipticity limits still in some cases approach a few times $10^{-6},$
the rough maximum currently expected from normal neutron
stars~\cite{JohnsonMcDaniel2013, Baiko2018}.

We are working on code to more efficiently handle high frequencies and long
spin-down ages.
With these improvements and ever improving strain noise from Advanced LIGO, the
prospects for continuous \ac{GW} detection will improve.

\acknowledgments

This research has made use of data, software and/or web tools obtained from the
Gravitational Wave Open Science Center (https://www.gw-openscience.org), a
service of LIGO Laboratory, the LIGO Scientific Collaboration and the Virgo
Collaboration. LIGO is funded by the U.S. National Science Foundation. Virgo is
funded by the French Centre National de Recherche Scientifique (CNRS), the
Italian Istituto Nazionale della Fisica Nucleare (INFN) and the Dutch Nikhef,
with contributions by Polish and Hungarian institutes.  This research was 
supported in part by NSF grants PHY-1604244, DMS-1620366, and PHY-1912419
to the University of California at San Diego; and by PHY-1912625 to Texas Tech
University.
The authors acknowledge computational resources provided by the High Performance
Computing Center (HPCC) of Texas Tech University at Lubbock
(\texttt{http://www.depts.ttu.edu/hpcc/}).

\bibliography{paper}

\begin{thebibliography}{59}
\bibitem [{EPA()}]{EPAPS}%
  \BibitemOpen
  \href@noop {} {}\bibinfo {note} {PRD will insert URL here}\BibitemShut
  {NoStop}%
\end{thebibliography}

\end{document}